\documentclass[prd,twocolumn,aps,preprintnumbers,amsmath,amssymb,nofootinbib]{revtex4}

\usepackage{graphicx}
\usepackage{color}
\usepackage{soul}
\usepackage{latexsym}
\usepackage{epsf}
\usepackage{amsmath}
\usepackage{epsfig}
\usepackage{hyperref}
\usepackage{natbib}

\def\lsim{\mathrel{\rlap{\lower4pt\hbox{\hskip1pt$\sim$}}
    \raise1pt\hbox{$<$}}}         
\def\gsim{\mathrel{\rlap{\lower4pt\hbox{\hskip1pt$\sim$}}
    \raise1pt\hbox{$>$}}}
\def\fun#1#2{\lower3.6pt\vbox{\baselineskip0pt\lineskip.9pt
  \ialign{$\mathsurround=0pt#1\hfil##\hfil$\crcr#2\crcr\sim\crcr}}}

\input epsf

\def\beq{\begin{equation}}
\def\eeq{\end{equation}}
\def\bea{\begin{eqnarray}}
\def\eea{\end{eqnarray}}

\begin{document}

\title{Breit-Wigner Enhancement Considering the Dark Matter Kinetic Decoupling}
\author{Xiao-Jun Bi}
\author{Peng-Fei Yin}
\author{Qiang Yuan}

\affiliation{ $^1$Laboratory of Particle Astrophysics, Institute
of High Energy Physics, Chinese Academy of Sciences, Beijing
100049, P. R. China }

\begin{abstract}

In the paper we study the Breit-Wigner enhancement of dark matter
(DM) annihilation considering the kinetic decoupling in the
evolution of DM freeze-out at the early universe. Since the DM
temperature decreases much faster (as $1/R^2$) after kinetic
decoupling than that in kinetic equilibrium (as $1/R$) we find the
Breit-Wigner enhancement of DM annihilation rate after the kinetic
decoupling will affect the DM relic density significantly.
Focusing on the model parameters that trying to explain the
anomalous cosmic positron/electron excesses observed by
PAMELA/Fermi/ATIC we find the elastic scattering $Xf\to Xf$ is not
efficient to keep dark matter in kinetic equilibrium, and the
kinetic decoupling temperature $T_{kd}$ is comparable to the
chemical decoupling temperature $T_f\sim O(10) GeV$. The reduction
of the relic density after $T_{kd}$ is significant and leads to a
limited enhancement factor $\sim O(10^2)$. Therefore it is
difficult to explain the anomalous positron/electron excesses in
cosmic rays by DM annihilation and give the correct DM relic
density simultaneously in the minimal Breit-Wigner enhancement
model.

\end{abstract}
\maketitle

\section{Introduction}

The recent cosmic ray observations by PAMELA\cite{pamela-e},
ATIC\cite{atic} and Fermi\cite{fermi} have all reported an excess
of positrons and electrons from $\sim 10$ GeV up to $\sim 1$ TeV.
These anomalies have stimulated a lot of interests, especially
these excesses may be attributed to the signals of dark matter
annihilation in the Galaxy. If these extra positrons/electrons are
indeed from DM annihilation, it requires definite properties of
DM. For example, DM should annihilate into lepton final states
dominantly and should have a much larger annihilation cross
section ($\langle \sigma v \rangle\sim 10^{23}$cm$^3$s$^{-1}$)
than the natural value ($\langle \sigma v \rangle\sim
10^{-26}$cm$^3$s$^{-1}$) at freeze out
\cite{Cirelli:2008pk,Yin:2008bs,Liu:2009sq}. The annihilation
cross section at freeze out determines the DM relic density if DM
is generated thermally at the early universe.

In general, the DM annihilation cross section $\langle \sigma v
\rangle$ depends on the averaged velocity of DM. For example, in
the usual weakly interacting massive particle (WIMP) scenario,
$\langle \sigma v \rangle$ can be expanded to a form of
$a+b\langle v^2 \rangle+O(v^4)$ at the non-relativistic limit
\cite{early}. If the annihilation process is s-wave dominant,
$\langle \sigma v \rangle$ is a constant. For the p-wave
annihilation, $\langle \sigma v \rangle$ is proportional to
$\langle v^2 \rangle$. Therefore the DM annihilation by
p-wave is suppressed today than the decoupling time since the WIMP
usually has a velocity of $v\sim 10^{-1}$ at the freeze-out epoch
and cools when universe expands. The DM velocity near the solar
system is $v \sim 10^{-3}$, much smaller than that at the
decoupling epoch.

However, as indicated by the PAMELA, ATIC and Fermi data, we
actually ask for a much larger annihilation cross section today to
account for the excesses than that at the early universe.
Contrary to the analysis before for the p-wave annihilation we
require an annihilation form $\langle \sigma v \rangle$ depends on
$\sim 1/v^n$. This form leads to a large annihilation cross
section today with low DM velocity and explains the cosmic
positron anomaly and relic density simultaneously. Some mechanisms
are soon proposed to achieve this aim after these results
published, such as the Sommerfeld enhancement
\cite{sommerfeld,ArkaniHamed:2008qn} and the Breit-Wigner
enhancement \cite{Griest:1990kh,Gondolo:1990dk,Ibe:2008ye,Guo:2009aj,Bi:2009uj}.

For the Sommerfeld enhancement, a new light mediator with mass of
O(GeV) is introduced, and provide an enhancement factor of $S\sim
\pi \alpha_X / v$ ($\alpha_X$ is coupling constant between DM and
mediator). For the Breit-Wigner enhancement, the DM annihilates
via a narrow resonance, and an enhance factor of $S\sim
max[\delta, \gamma]^{-1}/O(10)$ can be obtained \cite{Ibe:2008ye}
($\delta$, $\gamma$ are defined as $\delta=(4m^2-M^2)/4m^2$ and
$\gamma=\Gamma/M$ respectively, where m is the mass of DM, M and
$\Gamma$ are the mass and decay width of the resonance
respectively). One can achieve correct enhancement factor
$S=\langle \sigma v \rangle_{_{T=0}}/\langle \sigma v
\rangle_{_{T\sim T_{f}}}$ ($T_{f}$ is the temperature of chemical
decoupling) by adjusting the parameters appropriately.

It seems that the enhancement should not be important in the early
universe when the velocity of DM is $\sim O(10^{-1})$, and the
enhancement factor is only $S\sim O(1)$. However, some recent
studies showed that such effects are not negligible even at the
freeze-out epoch
\cite{Dent:2009bv,Zavala:2009mi,Feng:2009hw,Feng:2010zp},
especially for the Sommerfeld enhancement.
The Ref. \cite{Feng:2009hw,Feng:2010zp} pointed out that it may be
difficult to achieve the required enhancement factor in the
minimal Sommerfeld models considering the effect at the early
universe.

In this work, we will give a careful inspection on the
Breit-Wigner mechanism at the DM freeze-out process. For the
Breit-Wigner mechanism, the DM annihilation continues after the
chemical decoupling until the DM velocity drops below the cut-off
scale. Therefore the relic density is determined by the cut-off
scale related to $\delta$ and $\gamma$ \cite{Bi:2009uj}. In the
work we will show another important factor in determining the
relic density, \textit{i.e.} the kinetic decoupling process
\cite{Chen:2001jz,Hofmann:2001bi,Bringmann:2006mu}.

After the chemical decoupling at $x \sim 20$ ($x$ represents the
temperature of the universe which is defined as $x=m/T$), the DM
particle is still kept in kinetic equilibrium via the scattering
with the hot bath. When such scattering is not efficient to keep
DM in kinetic equilibrium, the DM momentum is red-shifted with the
scale factor $R$, which leads to a rapid decrease of DM
temperature as $T_X\sim R^{-2}$ rather than $T_X \sim R^{-1}$ at
the kinetic equilibrium epoch
\cite{Chen:2001jz,Hofmann:2001bi,Bringmann:2006mu}. Therefore,
after the kinetic decoupling $\langle \sigma v \rangle$ increases
quickly and then reduces the abundances of DM more efficiently.
Taking this effect into account we find the Breit-Wigner mechanism
is hard to provide a self-consistent explanation for both the DM
relic density and the positron anomaly today.

This paper is organized as following. In Section II,  we briefly
describe the Breit-Wigner enhancement mechanism at the DM
freeze-out epoch. In Section III, we discuss the kinetic
decoupling process. We will calculate the kinetic decoupling
temperature and the DM relic density including such effect. In
Section IV, we investigate the enhancement factor required by the
cosmic positron measurements. We will study the parameter space
and discuss whether there exists such parameters to explain all
the observations. Finally we give our conclusions and discussions
in Section V.

\section{the Breit-Wigner enhancement}
\label{sec: BWE}

In Ref. \cite{Ibe:2008ye}, the DM annihilation process is assumed
through $X\bar{X} \to R\to f\bar{f}$, where $R$ is a narrow
resonance with mass $M=\sqrt{4m^2 (1-\delta)}$ and  decay width
$\Gamma=M\gamma$ with $|\delta|, \gamma \ll 1$. For a scalar
resonance, the annihilation cross section is given as,
\begin{equation}
\sigma=\frac{16\pi}{M^2 \bar{\beta}_i \beta_i} \frac{\gamma^2}{(\delta+v^2/4)^2+\gamma^2}B_i B_f ,
\end{equation}
where $\bar{\beta}_i$ and $\beta_i$ are defined as
$\sqrt{1-4m^2/M^2}$ and $1-4m^2/s$ respectively, $s$ is given by
$s=(p_1+p_2)^2$, $B_i$ and $B_f$ denote the branching fractions of
the resonance into initial and final states respectively, $v$ is
the relative velocity of two initial particles. For $\delta >0$,
there exists an un-physical pole, but $B_i/\bar{\beta}_i$ is well
defined. For simplicity, we parameterize the cross section as
\cite{Ibe:2008ye}
\begin{equation}
\sigma v=\sigma_{_0}\frac{\delta^2+\gamma^2}{(\delta+z)^2+\gamma^2}.
\label{sigmavo}
\end{equation}
Here $\sigma_{_0}=\sigma v|_{_{T=0}}$ means the cross section at
zero temperature limit which is velocity independent, and is set
as a free parameter in our work \footnote{ For the scalar
resonance discussed above, $\sigma_{_0}$ is $\frac{32\pi B_i
B_f}{M^2 \bar{\beta}_i} \frac{\gamma^2}{\delta^2+\gamma^2}$. For
the $Z'$ model in Ref. \cite{Bi:2009uj} , $\sigma_{_0}$ denotes
$\frac{a^2g'^4}{16\pi m^2} \frac{1}{\delta^2+\gamma^2}$.
$\sigma_{_0}$ is a combination of $\delta$, $\gamma$ and other
parameters determined by the detailed model. It is indeed a free
parameter here. For more general discussions about the cross
section formula of DM annihilation via s-channel resonance, see
Ref. \cite{Backovic:2010ke} }. $z$ is defined in the form of
$s\equiv 4m^2(1+z)$ which equals $v^2/4$ in the
non-relativistic limit.

In order to calculate the DM relic density, it is necessary to
solve the Boltzmann equation \cite{early}
\begin{equation}
\frac{dY}{dx}=-\lambda' x^{-2} \langle \sigma v \rangle(Y^2-Y^2_{eq})
\label{Boeq}
\end{equation}
where $Y=n_\text {DM}/s$ is the DM number density normalized by
the entropy density $s$, $\lambda'$ is defined as
$\lambda'=\frac{s}{H}|_{_{x=1}}$. The entropy density $s(x)$ and
the Universe expansion rate $H(x)$ of the universe are given by
\begin{equation}
s(x) =  \frac{2 \pi^2 g_{*S}}{45} \frac{m^3}{x^3} \;\;, \;
H(x) = \sqrt{\frac{4 \pi^3 g_*}{45m_{pl}^2}} \frac{m^2}{x^2}
\;,
\end{equation}
where $g_*$($g_i$) is the effective number of degrees of freedom
for radiations (DM), and $g_{*S}$ is the effective number of
degrees of freedom defined by the entropy density. The $\langle
\sigma v \rangle$ can be parameterized as $\langle \sigma v
\rangle=\sigma_{_0}x^{-n}$ and the chemical decoupling temperature
is obtained as \cite{early}
\begin{equation}
x_f\simeq ln \varepsilon -(n+1/2)ln (ln\varepsilon),
\label{exf}
\end{equation}
where $\varepsilon\equiv c(c+2)a\lambda$ ($c\sim 1$ is a
constant), $\lambda\equiv \lambda' \sigma_{_0}$, and
$a=0.145(g_i/g_*)$ is defined in the form of
$Y_{eq}=ax^{3/2}e^{-x}$ at low temperature. The final $Y$ as $x$
tends to $\infty$ could be obtained approximately as
$Y_{\infty}\simeq (n+1)x_f^{n+1}/\lambda$ , and then the relic
density $\Omega_X h^2=2.74 \times 10^8 \frac{m}{GeV} Y_{\infty}$.

In Ref. \cite{Ibe:2008ye}, after parameterizing $\langle \sigma v
\rangle$ for $\delta>0$, the Boltzmann equation could be rewritten as,
\begin{equation}
\frac{dY}{dx}=-\frac{\lambda}{x^2}  \frac{\delta^2+\gamma^2}{(\delta+\xi x^{-1})^2+\gamma^2} (Y^2-Y^2_{eq}),
\label{mBE1}
\end{equation}
where $\xi\approx 1/\sqrt{2}$ is a constant (in fact, there is an
assumption here that $x\sim v^{-2}$ or DM stays in kinetic
equilibrium until very low temperature in Eq. (\ref{mBE1})). For
the Breit-Wigner enhancement, the freeze-out process begins at
$\tilde{x}_f\sim O(10)$ \footnote{The $\tilde{x}_f$ could be
achieved approximately by setting $\lambda \to \lambda
(\delta^2+\gamma^2)/\xi$ and $n\to -2$ in the Eq. (\ref{exf}).},
and continues until the temperature of $x_b\simeq max[\delta,
\gamma]^{-1}$ when the DM annihilation cross section does not
increase with the universe cooling. The final value of $Y$ is
$Y_{\infty}\simeq x_b/\lambda$. In the ordinary S-wave
non-resonant annihilation scenario with $\langle \sigma v
\rangle=constant$, one could obtain $Y_{\infty}\simeq
x_{f0}/\lambda_0$, where $x_{f0}\sim 20$, $\lambda_0\simeq
\lambda'\times 10^{-9}GeV^{-2}$. Then the enhancement factor is
achieved as $S\simeq x_b/x_{f0}\simeq max[\delta,
\gamma]^{-1}/O(10)$ \cite{Ibe:2008ye}. The Breit-Wigner
enhancement has been used to explain the anomalous positron
excesses which require an enhancement factor of $\sim O(10^3)$.

\section{Kinetic Decoupling of DM particles}

\begin{figure*}[t]
\begin{tabular}{ccc}
\includegraphics[totalheight=1.5in,clip=true]{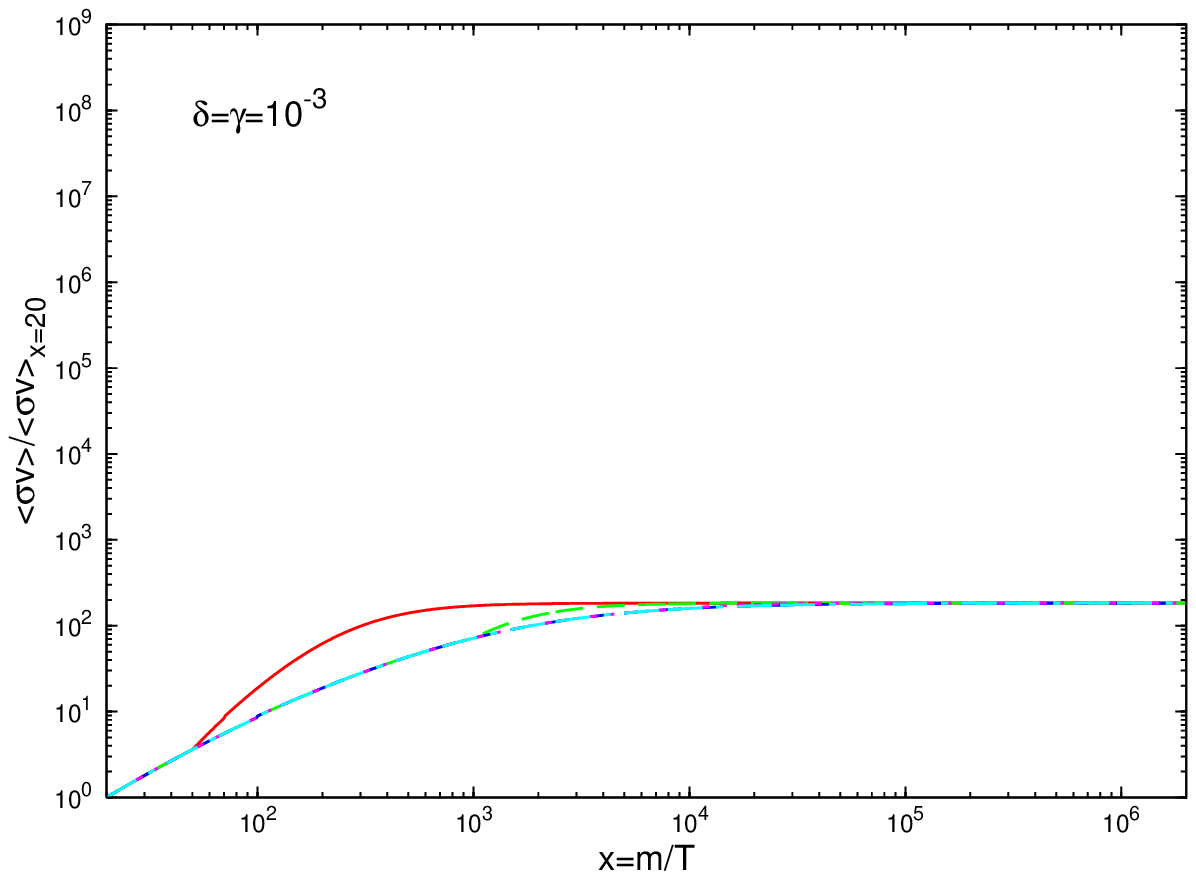} &
\includegraphics[totalheight=1.5in,clip=ture]{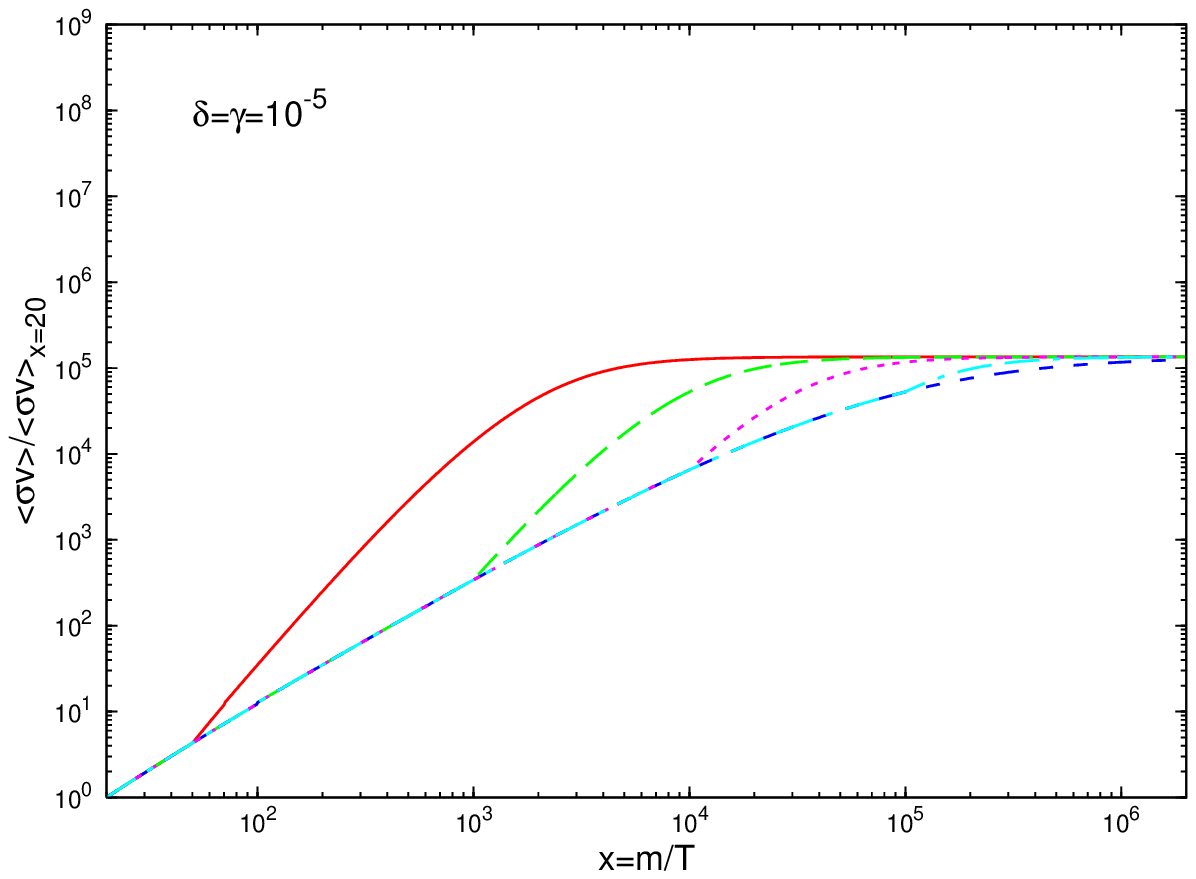} &
\includegraphics[totalheight=1.5in,clip=ture]{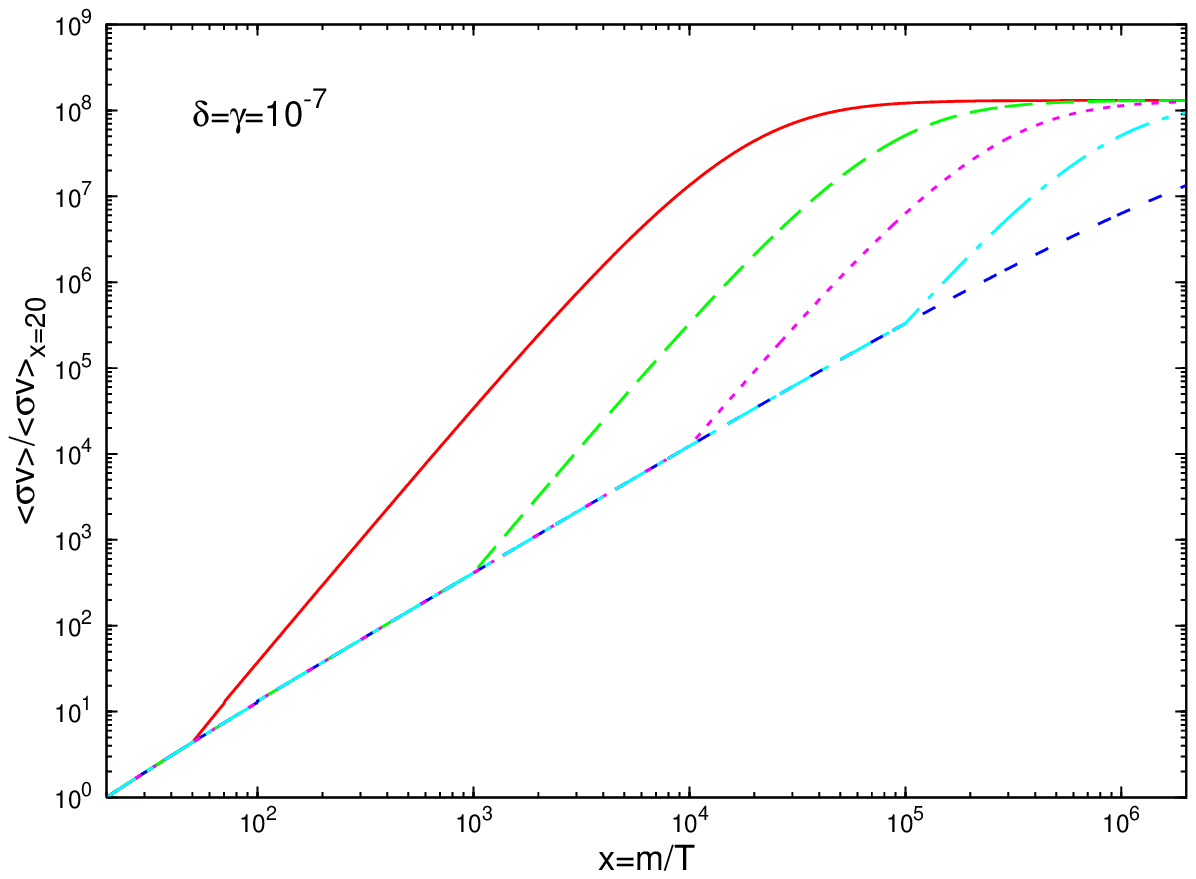}
\end{tabular}
\caption{The Breit-Wigner enhanced relative cross section $\langle
\sigma v \rangle/\langle \sigma v \rangle|_{_x=20}$ as a function
of $x$.  The curves in the figures from left to right denote
$x_{_{kd}}=50, 10^3, 10^4, 10^5, \infty$ respectively. The model
parameters in the figures from left to right are set as
$\gamma=\delta=10^{-2}, 10^{-5}, 10^{-7}$ respectively.}
\label{Fig:enhance}
\end{figure*}

In the early universe, the DM production and annihilation
processes $X\bar{X}\rightleftharpoons f\bar{f}$ are efficient to
keep DM particles in chemical and kinetic equilibrium. After
chemical decoupling at $T_f$ DM may keep in kinetic equilibrium by
momentum exchange with the hot bath of the standard model
particles via the t-channel scattering $Xf\to Xf$, until the
temperature decreases to the kinetic decoupling temperature
$T_{kd}$.

Before kinetic decoupling, the DM has the same temperature as the
thermal bath. After kinetic decoupling, the temperature of DM
$T_X$ decreases as $1/R^2$, while the the temperature of thermal
radiation still decrease as $1/R$. So the $T_X$ could be
determined as \cite{Chen:2001jz,Hofmann:2001bi,Bringmann:2006mu}
\begin{eqnarray}
\left\{\begin{array}{cc} T_X=T, \;\;\; T_X> T_{kd}\\
\;\;\;\;T_X=T^2/T_{kd}, \;\;\; T_X \leq  T_{kd}
\end{array}\right..
\label{txtkd}
\end{eqnarray}

Since $T_X$ is different from $T$ one can define a parameter
$x_{_X}$ related to DM temperature $T_X$ as
\begin{equation}
x_{_X}=\frac{m}{T_X}=\frac{2}{v_0^2},
\end{equation}
where $v_{_0}$ is the most probable velocity of DM. The $\langle
\sigma v \rangle$ is a function of $x_{_X}$ which is given by
\cite{Gondolo:1990dk}
\begin{equation}
\langle \sigma v \rangle = \frac{1}{n_{EQ}^2} \frac{m}{64 \pi^4 x_{_X}}
\int_{4 m^2}^{\infty} \hat{\sigma}(s) K_1(\frac{x_{_X}
\sqrt{s}}{m}) d s \;, \label{cross}
\end{equation}
with
\begin{eqnarray}
n_{EQ} & = & \frac{g_i}{2 \pi^2} \frac{m^3}{x_{_X}} K_2(x_{_X}) \; , \label{n} \\
\hat{\sigma}(s) & = & 4 E_1 E_2 \sigma v  g_i^2 \sqrt{1-\frac{4m^2}{s}} \; , \label{sigma}
\end{eqnarray}
where $K_1(x)$ and $K_2(x)$ are the modified Bessel functions of the first and second type respectively.

After kinetic decoupling, the temperature of DM decreases rapidly,
and the Breit-Wigner enhancement increases significantly. In the
Fig. \ref{Fig:enhance}, we show the enhancement factor of $\langle
\sigma v \rangle/\langle \sigma v \rangle|_{_x=20}$ for
$x_{_{kd}}=50, 10^3, 10^4, 10^5$ respectively. We also give the
results in the limit of $x_{_{kd}}=\infty$ which denotes no
kinetic decoupling. From Fig. \ref{Fig:enhance}, we can see the
$\langle \sigma v \rangle$ for $x_{_{kd}}=50$ increases more
quickly reaching the maximal value than the cases without kinetic
decoupling. On the other hand, for a large value of
$x_{_{kd}}=10^4, 10^5$, such effects are not very obvious. These
results could be understood easily from Eq. (\ref{sigmavo}) by
assuming $\langle \sigma v \rangle \sim \sigma v|_{z\to v_0^2} $
roughly. When $x>x_{_{kd}}$ and $v_0^2\gg \delta, \gamma$,
$\langle \sigma v \rangle$ increase as $x^{2}/x_{_{kd}}$ rather
than $x$, and reaches $\sigma_0$ more quickly for small
$x_{_{kd}}$.

\begin{figure}[t]
\includegraphics[totalheight=2.1in]{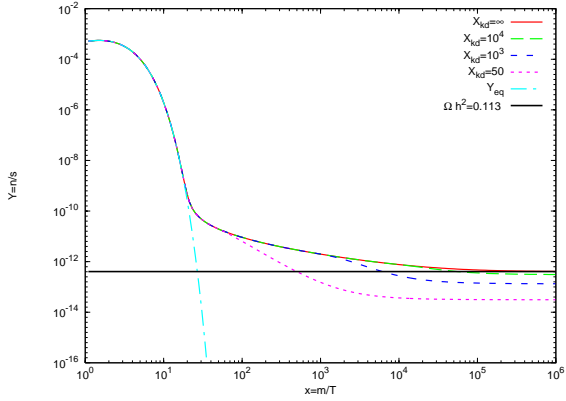}
\caption{The evolution of DM abundance $Y$ as a function of $x$. The $x_{_{kd}}$ are taken as $50, 10^3, 10^4, \infty$ respectively.}
\label{Fig:relic_d}
\end{figure}

The Fig. \ref{Fig:relic_d} shows the effects of kinetic decoupling
in the calculation of the relic density. After kinetic decoupling,
the annihilation of DM becomes more significantly, and reduce the
relic density more efficiently. If kinetic decoupling is very late
$x_{_{kd}}\gg\tilde{x}_f$, for example $x_{_{kd}}=10^4$, such
effect is not very important compared with the case without
kinetic decoupling. However, if the kinetic decoupling occurs at
nearly the same epoch as the chemical decoupling, the efficient
annihilation would reduce DM relic density by about one order of
magnitude. Therefore the kinetic decoupling temperature $T_{kd}$
is a very important parameter in the calculation of the DM relic
density.

The kinetic decoupling temperature $T_{kd}$ can be determined
using the method in Refs. \cite{Feng:2010zp,Hofmann:2001bi}. If
the momentum transfer rate drops below the expansion rate, the DM
decouples from the kinetic equilibrium with the radiation
background. Therefore, the $T_{kd}$ can be determined
approximately by the relation of $\Gamma_k(T_{kd})=H(T_{kd})$. The
momentum transfer rate is defined as Ref.
\cite{Feng:2010zp,Hofmann:2001bi}
\begin{equation}
\Gamma_k\sim n_r \langle \sigma v \rangle_s \frac{T}{m}
\end{equation}
where $n_r$ is the number density of massless fermions with
$n_r=\frac{3}{4}\cdot \frac{1.202}{\pi^2}g_f T^3$, $\langle \sigma
v \rangle_k$ is the thermally averaged cross section for the
scattering process $Xf\to Xf$. Note that there exists a factor of
$T/m$ in the above formula, which reflects the approximate
momentum transfer at each collision.

Since the elastic scattering $Xf\to Xf$ via t-channel is
suppressed by the propagator of $R$ with $1/(t-M^2)^2\sim 1/M^4$, the cross section of this process is much
smaller than the annihilation cross section. The explicit formula
of the cross section for $Xf\to Xf$ depends on the model details.
An approximate cross section of $Xf\to Xf$ is related to
$X\bar{X}\to f\bar{f}$ as
\begin{equation}
\sigma v_s \sim  a_s \sigma_{_0} (\delta^2+\gamma^2)\frac{T^2}{m^2}\;\; ,
\label{sigmak}
\end{equation}
where $a_s \leq O(1)$ is a constant determined by the form of the
interaction (for more details, see the appendix). Then we can
estimate $T_{kd}$ by setting $\Gamma_k(T_{kd})= H(T_{kd})$. Then
we get
\begin{eqnarray}
T_{kd} & \sim & 2.0\left[\frac{\sqrt{g_{_*}} m^3}{g_f a_s \sigma_0 (\delta^2+\gamma^2)m_{_{pl}}}\right]^{\frac{1}{4}} \nonumber \\
 & \sim & 30. GeV \left[ \frac{1}{a_s}\right]^{\frac{1}{4}} \left[ \frac{10^{-6} GeV^{-2}}{\sigma_{_0}}\right]^{\frac{1}{4}} \left[\frac{\; 10^{^{-9}}}{\delta^2+\gamma^2}\right]^{\frac{1}{4}} \nonumber \\
 &\times& \left[\frac{4}{g_f}\right]^{-\frac{1}{4}} \left[\frac{g_{_*}}{100}\right]^{\frac{1}{8}} \left[\frac{m}{1TeV}\right]^{\frac{3}{4}} \;\; . \label{tkd}
\end{eqnarray}
From above estimation, we can see the typical $T_{kd}$ in the Breit-Weigner enhancement model is O(10)GeV, which is much larger than that in the ordinary WIMP model. For example, the $T_{kd}$ for neutralino in the SUSY model is only O(10)MeV \cite{Chen:2001jz,Hofmann:2001bi}.

If $T_{kd}$ in Eq. (\ref{tkd}) is larger than the DM freeze-out
temprature\footnote{Here we define $\tilde{x}_f$ as the time when
$n_{_{DM}}(\tilde{x}_f)=10n^{eq}_{_{DM}}(\tilde{x}_f)$.} $T_f\sim m/\tilde{x}_f \sim
m/20$, it means the elastic scattering becomes unimportant before
the chemical decoupling. However, the DM particles are still kept
in thermal equilibrium by the annihilation process $f\bar{f}\to
XX$. Therefore $T_{_{kd}}$ should be defined as $min(T_f,
T_{kd}')$, where $T_{kd}'$ is determined by the elastic scattering
as given in Eq. (\ref{tkd}).

\begin{figure*}[t]
\begin{tabular}{cc}
\includegraphics[totalheight=2.3in,clip=true]{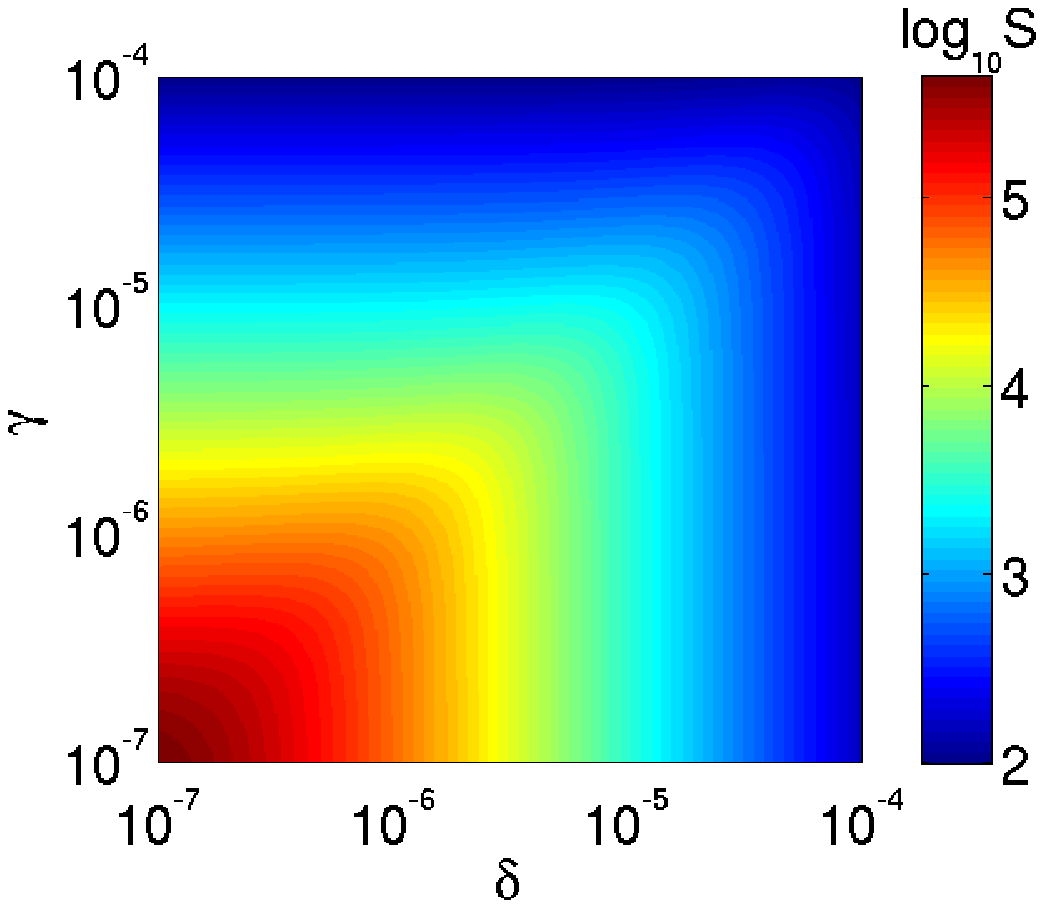} &
\includegraphics[totalheight=2.3in,clip=ture]{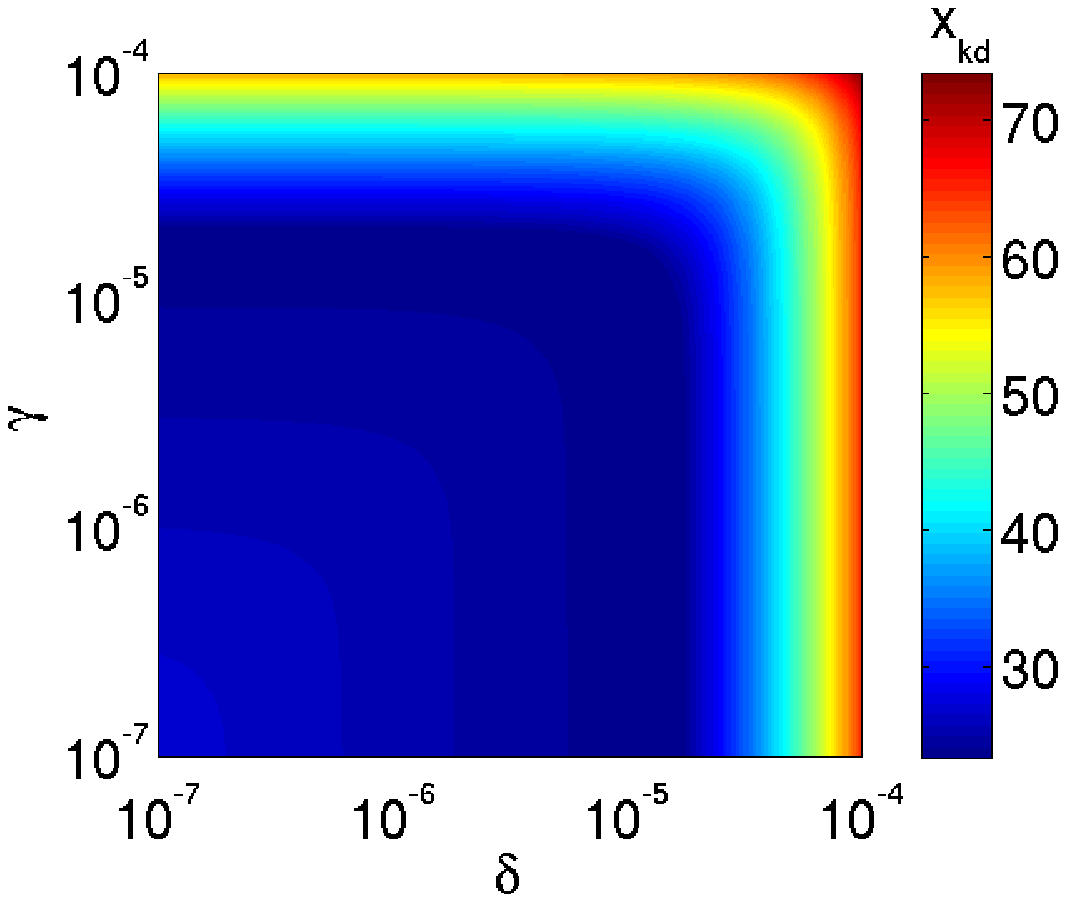}
\end{tabular}
\caption{Numerical illustration of the DM annihilation enhancement
factor $\sigma_0/\sigma_{nature}$ without considering kinetic decoupling effect (left) and the corresponding $x_{kd}$ (right) on the
$\gamma-\delta$ plane.}
\label{Fig:boostnotkd}
\end{figure*}

For a more precise calculation, one need to derive the DM temperature $T_{_X}(T)$ from the Boltzmann equation
\begin{equation}
L[f]=C[f] \;\;,
\end{equation}
where the $L$ and $C$ are Liouville operator and collision operator for the scattering process respectively. A general relation between the $T_{X}$ and $T$ has been provided by Ref. \cite{Bringmann:2006mu}. In our work, we still adopt the simple relation between the $T_{X}$ and $T$ as Eq. (\ref{txtkd}), and use the formulae in Ref. \cite{Bringmann:2006mu} to calculate the $T_{kd}'$. We take a $Z'$ model with $m_{DM}=1$TeV as an example, but our results can be extended to other models (for more details, see the appendix). From our calculations, we find that for the typical
parameters used to explain the PAMELA/Fermi/ATIC results $
x_{_{kd}}=m/T_{_{kd}}$ is not far from $\tilde{x}_f \sim O(10)$.
To show this point explicitly, we give the boost factor
$S=\sigma_{_0}/\sigma_{_{nature}}$ (left) and $x_{_{kd}}$ (right)
for different $\delta$ and $\gamma$ in Fig. \ref{Fig:boostnotkd}.
Here $\sigma_{_{nature}}=3\times 10^{-26}$ cm$^3$s$^{-1}$ is the
so called `natural' value of DM annihilation cross section
predicted by the WIMP models to generate correct relic density. In
the left plot, we require each point in the parameter space
producing the correct relic density without kinetic decoupling
effect, and determine the corresponding $\sigma_{_0}$. Then we use
these $\delta$, $\gamma$ and $\sigma_{_0}$ to calculate
$x_{_{kd}}$. We find in the parameter space favored by the
PAMELA/Fermi/ATIC results with $S\sim O(10^3)$, $ x_{_{kd}}$ is
similar as $\tilde{x}_f$. It means the kinetic decoupling effect
should be important in the early universe when determining the DM
relic density. Therefore it should be considered carefully in the
explanation of the anomalous cosmic positron flux.

\section{the enhancement factor for anomalous positron/electron flux}
\label{sec: bfp}

In this section, we calculate the enhancement factor by the
Breit-Wigner resonance in the Galaxy today considering the kinetic
decoupling.
Here we define the enhancement factor as
\begin{equation}
S=\sigma_{_G}/\sigma_{_{nature}} \;\; ,
\label{enhancefactor}
\end{equation}
where $\sigma_{_G}$ denotes the $\langle \sigma v \rangle$ of DM
with the most probable velocity $v_{_G}\sim 10^{-3}$ in the
Galaxy. This definition is different from the earlier form
$s=\sigma_{_0}/\sigma_{_{nature}}$ \cite{Ibe:2008ye,Guo:2009aj} as
the DM velocity is not zero today. We will see such difference is
important.

\begin{figure}[t]
\includegraphics[totalheight=2.5in]{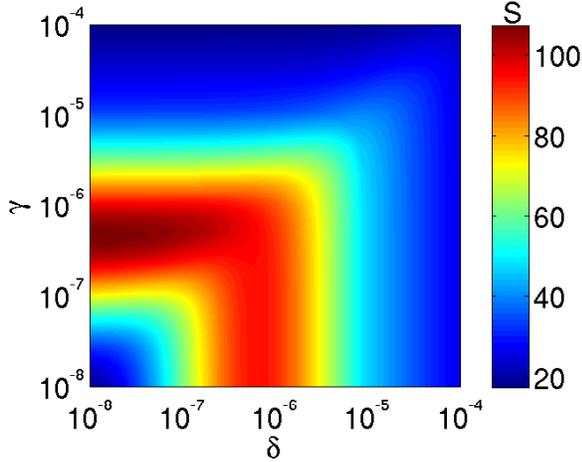}
\caption{Numerical illustration of the enhance factor of $S =
\sigma_{_{G}}/\sigma_{_{nature}}$ on the
$\gamma-\delta $ plane for the $\delta>0$ case. }
\label{Fig:boost1}
\end{figure}

\begin{figure}[t]
\includegraphics[totalheight=2.5in]{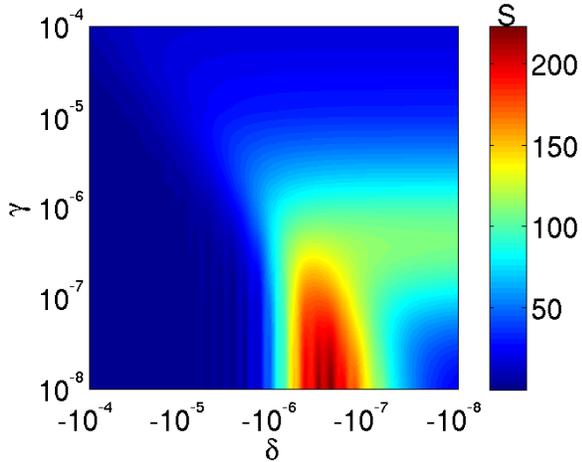}
\caption{The same as Fig. \ref{Fig:boost1} but for $\delta<0$. }
\label{Fig:boost2}
\end{figure}

We give the numerical results of $S$ in Fig. \ref{Fig:boost1} and
Fig. \ref{Fig:boost2} for the cases of  $\delta>0$ and $\delta<0$
respectively. For each point in the two figures, $\sigma_0$ has
been adjusted to produce the correct relic density. The maximum
value of $S$ is only $O(10^2)$ with $\delta, \gamma \sim
O(10^{-6})$ . From these results, we find the Breit-Wigner
enhancement is difficult to provide large enough boost factor to
explain the anomalous positron excesses after taking into account
the kinetic decoupling effect.

To check this result analytically, one would turn to the
discussion in the last paragraph of Sec. \ref{sec: BWE}
\cite{Ibe:2008ye}. After the kinetic decoupling, the $\xi x^{-1}$
in Eq. (\ref{mBE1}) should be modified by $\xi x_{kd} x^{-2}$. The
DM annihilation would continue to the temperature of $x_b \sim
max[\delta, \gamma]^{-\frac{1}{2}}\cdot \sqrt{x_{kd}}$. One can
also obtain an enhancement factor as $S\sim x_b/x_{f0}\sim
max[\delta, \gamma]^{-\frac{1}{2}}\cdot \sqrt{x_{kd}}/x_{f0}$. It
seems we could still achieve a required boost factor by taking
some smaller parameters such as $(\delta,\gamma) \sim
O(10^{-(6\sim 8)})$. However, this is not the case. In fact, one
can indeed obtain an arbitrary value of
$\sigma_{_0}/\sigma_{_{nature}}$ by setting the $\delta$ and
$\gamma$ tiny enough as discussed above. But the factor of
$\sigma_{_{G}}/\sigma_{_{nature}}$ is different with that in the vanishing DM
velocity limit. From the Eq. (\ref{sigmavo}) we can see, for the
parameters of $\delta \simeq \gamma \geq O(10^{-5})$, these two
factors are equal, because the $\langle \sigma v \rangle$ always
reaches its maximum value $\sigma_{_0}$ when the DM velocity
decreases to $v_{_G}\sim z\sim 10^{-3}<
max[\delta,\gamma]^{\frac{1}{2}}$. However, for $\delta \simeq
\gamma < O(10^{-6})$, $\sigma_{_G} \sim
\sigma_{_0}(max[\delta,\gamma]/v_{_G}^2)^2$ is always smaller than
$\sigma_{_0}$.

To understand the maximum value of
$\sigma_{_{G}}/\sigma_{_{nature}}$ and the corresponding
parameters in Fig. \ref{Fig:boost1} and Fig. \ref{Fig:boost2}, we
can also use Eq. (\ref{sigmavo}) as a roughly estimation. By
setting $z\sim v_{_G}^2\sim 10^{-6}$ and
$\sigma_{_0}/\sigma_{_{nature}} \sim  max[\delta,
\gamma]^{-\frac{1}{2}}\cdot \sqrt{x_{kd}}/x_{f0}$, we could obtain
$S$
\begin{equation}
S\sim \frac{max[\delta, \gamma]^{\frac{3}{2}}}{max[\delta+v_{G}^2, \gamma]^2}\cdot \frac{\sqrt{x_{kd}}}{x_{f0}}  \;\; .
\label{enhancefactorsim}
\end{equation}
From this rough estimation, we can see for the $\delta > 0$ case,
there actually exists a maximum value of $S$ around $max[\delta, \gamma]\sim
v_G^2\sim O(10^{-6})$ as shown in the Fig. \ref{Fig:boost1}. On
the other hand, for the $\delta < 0$, when $\gamma \gg \delta+v_{_G}^2 \to 0
$, the $S$ might be larger than the case of $\delta>0$. It
means at the physical pole resonance, if the annihilations in the
galaxy occur accurately with $v^2/4-\delta \to 0$, the cross
section could be very large. However, considering the dispersion
of the DM velocity our numerical results show the enhancement
factor can not be very large either.

\section{Conclusion and Discussion}
\label{sec:con}

In this work, we study the Breit-Wigner enhancement for DM
annihilation taking the kinetic decoupling effect at the early
universe into account. We find if the kinetic decoupling occurs at
nearly the same epoch as the chemical decoupling, the DM
annihilation process becomes very important and reduces the DM
relic density significantly.
Requiring the model gives correct relic density we find there is
no parameter space that can give an annihilation cross section
today large enough to explain the anomalous cosmic
positron/electron excesses at PAMELA/ATIC/Fermi.

The main point here is the elastic scattering between DM and
massless fermions $Xf\to Xf$ is not efficient to maintain DM in
thermal equilibrium. The kinetic decoupling occurs at high
temperature $\sim T_f$. The DM temperature would decrease as $\sim
T^2/T_{f}$ after kinetic decoupling, and reaches a very small
value before the structure formation. For typical WIMP such as
neutralino, the typical damping mass is $\sim 10^{-6} M_\odot
(m/100GeV)^{-\frac{3}{2}} (T_{kd}/30MeV)^{-\frac{3}{2}}$
\cite{Bringmann:2006mu}. Therefor in the Breit-Wigner enhancement
with high $T_{kd}$, the damping mass might be much smaller than
the usual cold DM model. This kind of DM model may predict tiny DM
subhalo with $M_{_{sub}}\ll 10^{-6} M_\odot$ in the Galaxy.
The realistic impact for the structure formation in the Breit-Wigner
mechanism may need a careful study. This feature is possible to change the predictions for DM indirect
detection.

Finally we would like to point out that it is still possible to
explain the anomalous cosmic positron excesses in some non-minimal
Breit-Wigner models. The ideal here is adding some new interaction
process to keep DM in kinetic equilibrium till to a low
temperature. For example, the DM is slepton $\tilde{\tau}$ in the
hidden sector \cite{Feng:2009mn}. It might interact with the
hidden photon with a large coupling constant, or scatter with the
standard model leptons by exchanging hidden neutralino in
resonance. The hidden slepton annihilation to leptons could be
enhanced by a $Z'$ resonance in the $U(1)_{L_i-L_j}$ model
\cite{Bi:2009uj}. With this setting to enhance the scattering
process, it is possible to obtain a low $T_{kd}$, and recover all
the discussions in the earlier works about the Breit-Wigner
mechanism.

\acknowledgements We would like to thank Haibo Yu for helpful
discussions. This work is supported in part by the Natural
Sciences Foundation of China (No. 11075169), the 973 project under
the grant No. 2010CB833000 and the by the Chinese Academy of
Science under the grant No. KJCX2-EW-W01.

\begin{appendix}

\section{relation between cross sections of annihilation and elastic scattering processes }

In this appendix, we would give detailed discussions about the relation between the cross sections of annihilation $X\bar{X}\to f\bar{f}$ and elastic scattering $Xf \to Xf$.

We assume the effective interaction Lagrangian between two DM particles (X) and two leptons (f) as
\begin{equation}
g_A g_B R_0 \bar{X}\Gamma_X X \bar{f} \Gamma_f f
\end{equation}
where $g_A$ and $g_B$ are interaction couplings of $X\bar{X} R$ and $f\bar{f} R$ respectively, $\Gamma_X$ and $\Gamma_f$ are combines of Lorentz metrics determined by model, $R_0$ is the propagator of resonance $R$. The cross section of annihilation process is (we neglect SM fermion mass $m_{_f}$ here)
\begin{equation}
\frac{d \sigma}{d\Omega} = \frac{1}{64\pi^2 s} \sqrt{\frac{s}{s-4m^2}} g_A^2 g_B^2 |R_{0a}|^2 |M'_a|^2 \;\; ,
\label{sigmaann}
\end{equation}
where we define the squared transition matrix element is $g_A^2 g_B^2 |R_{0a}|^2 |M'_a|^2$, $s,t,u$ are usual Mandelstam variables. For annihilation process in the non-relativistic limit, $s$ is approximated as $4m^2+m^2v^2$. So we can achieve $\sigma_{_0} $ as
\begin{equation}
\sigma_{_0} = \frac{1}{32\pi m^2} \frac{g_A^2 g_B^2}{M^4(\delta^2+\gamma^2)}\int\frac{|M'_a|^2}{4\pi}d\Omega \;\; ,
\label{sigma0ann}
\end{equation}

For the cross section of elastic scattering, we find
\begin{equation}
\sigma v_{s} = \frac{1}{16\pi m^2} \frac{g_A^2 g_B^2}{M^4}\int\frac{|M'_s|^2}{4\pi}d\Omega \;\; ,
\label{sigmas}
\end{equation}
and the $\sigma v_{s}$ could be expressed by $\sigma_{_0}$ as
\begin{equation}
\sigma v_{s} = 2 \sigma_{_0} (\delta^2+\gamma^2)\frac{\int|M'_s|^2d\Omega}{\int|M'_a|^2d\Omega} \;\;.
\label{sigma0s}
\end{equation}
In general, the $|M'_a|^2$ and $|M'_s|^2$ are expressed by the four vector momentums of four particles. We define $p_1$, $p_2$ for two DM particles, and $k_1$, $k_2$ for two fermions. We need only calculate either  one of $|M'_a|^2$ and $|M'_s|^2$, and make some modifications to obtain the other one. In the calculation, we can neglect all the sub-leading terms which are proportional to $m_f^2$, $v^2$, and assumed the energy of fermion in the scattering process is $\omega=3T/2$.

For example, we can calculate the relation between $\sigma v_{s}$ and $\sigma_{_0}$ in a $Z'$ model with $\Gamma_X=\Gamma_f=\gamma^\mu$. The $|M'_a|^2$ is given by
\begin{eqnarray}
|M'_a|^2 &=& \frac{1}{4}\cdot 32 \cdot [ (k_1\cdot k_2)m^2+ (p_1\cdot k_1)(p_2 \cdot k_2)\nonumber \\
&+&(p_1\cdot k_2)(p_2 \cdot k_1) ]=32m^4 \;\; . \label{ma2}
\end{eqnarray}
Then we can achieve $\overline{|M'_s|^2}=8m^2\omega^2$ and $\sigma v_{s} = \frac{9}{8}\sigma_{_0} (\delta^2+\gamma^2)\frac{T^2}{m^2}$.

\section{calculation for the kinetic decoupling temperature}

In this appendix we show the calculation for the $T_{_{kd}}$ in a $Z'$ model. The detailed method is described in Ref. \cite{Bringmann:2006mu}, and can be extended to other models easily.

In general, the DM temperature $T_{_{X}}(T)$ can be derived by solving Boltzmann equation
\begin{equation}
T_{_X}=T\left[1-\frac{z^{\frac{1}{n+2}}}{n+2} exp[z] \Gamma[-(n+2)^{-1},z]\right]\;\; ,
\label{txex}
\end{equation}
where $z=\frac{a}{n+2}(\frac{T}{m})^{n+2}$. In the low (high) temperature limit $T\to 0$ ($T\to \infty$), the $T_{_X}$ has the same form $T_{_X}\to T^2/m$ ($T_{_X}\to T$)  as Eq. (\ref{txtkd}). Then the kinetic decoupling temperature can be obtained as
\begin{equation}
T_{_{kd}}=\left(\frac{T^2}{T_{_X}}\right)_{T\to 0}=m\left[\left(\frac{a}{n+2}\right)^{\frac{1}{n+2}} \Gamma\left[\frac{n+1}{n+2}\right]  \right]^{-1}\;\; ,
\label{tkdex}
\end{equation}
The parameters $a$ and $n$ are defined as follows. One need to expand the amplitude at $t=0$ and $s=m^2+2m\omega$
\begin{equation}
|M|^2=c_n\left(\frac{\omega}{m}\right)^n+O\left(\left(\frac{\omega}{m}^{n+1} \right)\right) \;\; ,
\end{equation}
The constant $a$ is given by
\begin{equation}
a=\sum_f \left( \frac{10}{(2\pi)^9 g_{*}} \right)^{1/2} g_f c_n N_{n+3}^\pm \frac{m_{pl}}{m} \;\; .
\end{equation}
The $N_{n+3}^\pm$ for fermion (plus sign) and scalar (minus sign) are given by
\begin{equation}
N_n^\pm=(1-p^\pm 2^{-n})(n+1)!\zeta(n+1)  \;\; ,
\end{equation}
where $p^+=1$ and $p^-=0$.
For a $Z'$ model with resonance mass $m_{Z'}=M \sim 2m $ described as
\begin{equation}
\frac{g_A g_B}{M^2} \bar{X}\gamma^\mu X \bar{f} \gamma_\mu f \;\; ,
\end{equation}
we can achieve the amplitude at zero momentum transfer as $g_A^2 g_B^2 |M'_s|^2_{t=0}/M^4=\frac{g_A^2 g_B^2}{2}(\frac{\omega}{m})^2$. Substituting $n=2$ and $c_2=g_A^2 g_B^2/2$ in the Eq. (\ref{tkdex}), we obtain $T_{_{kd}}$
\begin{equation}
T_{_{kd}}=1.326 \left[ \frac{\sqrt{g_*} m^5}{g_f c_2 M_{_{pl}}} \right]^{\frac{1}{4}} \;\; .
\label{tkdz}
\end{equation}
Here we assume the $Z'$ has the same couplings with the different leptons and sum the $g_f$ together. The thermal average annihilation cross section at low temperature can be written as $\sigma_0=c_2/8\pi m^2(\delta^2+\gamma^2)$, the Eq. (\ref{tkdz}) can be re-written as
 \begin{equation}
T_{_{kd}}=0.6 \left[ \frac{\sqrt{g_*} m^3}{g_f \sigma_0 (\delta^2+\gamma^2) M_{_{pl}}} \right]^{\frac{1}{4}} \;\; ,
\label{tkdz}
\end{equation}
which is smaller than the result from Eq. (\ref{tkd}) by a factor of O(1).

\end{appendix}

\end{document}